\documentclass[sigconf]{acmart}
\usepackage{subcaption}
\AtBeginDocument{%
  }

\copyrightyear{2026}
\acmYear{2026}
\setcopyright{cc}
\setcctype{by}
\acmConference[RecSys '26]{20th ACM Conference on Recommender Systems}{September 27-October 02, 2026}{Minneapolis, MN, USA}
\acmBooktitle{20th ACM Conference on Recommender Systems (RecSys '26), September 27-October 02, 2026, Minneapolis, MN, USA}
\acmDOI{10.1145/3773078.3831768}
\acmISBN{979-8-4007-2284-4/2026/09}
\begin{document}

\title{Structure-Preserving Projection for Mitigating Modality Bias in LLM-Based Sequential Recommendation}

\author{Tzu-Wei Chiu}
\orcid{0009-0003-1731-054X}
\authornote{Equal contribution.}
\affiliation{%
 \institution{National Taiwan University}
 \city{Taipei}
 \country{Taiwan}
}
\email{R12922A14@ntu.edu.tw}

\author{Song-Duo Ma}
\orcid{0009-0006-2000-3921}
\authornotemark[1]
\affiliation{%
 \institution{National Taiwan University}
 \city{Taipei}
 \country{Taiwan}
}
\email{R14944001@ntu.edu.tw}

\author{Hsin-Yu Lin}
\orcid{0009-0004-5885-9093}
\affiliation{%
 \institution{National Taiwan University}
 \city{Taipei}
 \country{Taiwan}
}
\email{R14922159@ntu.edu.tw}

\author{Pu-Jen Cheng}
\orcid{0000-0001-5892-0385}
\affiliation{%
 \institution{National Taiwan University}
 \city{Taipei}
 \country{Taiwan}}
\email{pjcheng@csie.ntu.edu.tw}

% \renewcommand{\shortauthors}{Chiu et al.}

\begin{abstract}
  Recent LLM-based recommenders integrate textual and collaborative signals by projecting collaborative embeddings into the embedding space of the LLM. However, this projection can introduce modality bias that distorts the underlying collaborative structure and limits the usefulness of projected embeddings. To address this issue, we propose a novel structure-preserving projection approach that maintains the relational geometry of collaborative embeddings through dedicated structure-preserving losses. Comprehensive experiments demonstrate that our approach consistently improves recommendation performance, providing a more reliable path for LLM-based recommendation.
  
\end{abstract}




\maketitle

\section{Introduction}

Sequential recommendation aims to predict users' next interaction item by modeling temporal patterns and behavioral transitions from historical sequences in large-scale platforms such as e-commerce, streaming, and social media \citep{fang2020deep,quadrana2018sequence,wang2019sequential}. Early approaches based on recurrent and self-attentive architectures, such as GRU4Rec \citep{hidasi2015session} and SASRec \citep{kang2018self}, established strong baselines by capturing sequential dependencies in collaborative signals. However, these models rely primarily on interaction patterns and co-occurrence statistics, limiting their ability to reason over heterogeneous signals such as item content and natural language descriptions.

Recently, LLM-based recommender systems have emerged as a promising direction for combining the language understanding capabilities of LLMs with collaborative signals derived from user interaction histories. Rather than relying solely on interaction patterns, these methods reformulate recommendation as a language modeling task and incorporate textual item information, such as titles or descriptions, into prediction. To further exploit collaborative signals learned by traditional recommenders, several studies project pretrained item embeddings into the LLM embedding space, allowing the model to condition on both textual and collaborative inputs. For example, CoLLM~\citep{zhang2025collm} formulates recommendation as a binary decision problem over textual item histories and collaborative representations; LLaRA~\citep{liao2024llara} performs next-title generation with a curriculum that gradually replaces placeholders with projected embeddings; and A-LLMRec~\citep{kim2024large} aligns collaborative representations with SBERT-based text before projection.

\begin{figure}[t]
    \centering
    \begin{minipage}[t]{1\columnwidth}
        \centering
        \includegraphics[width=\linewidth]{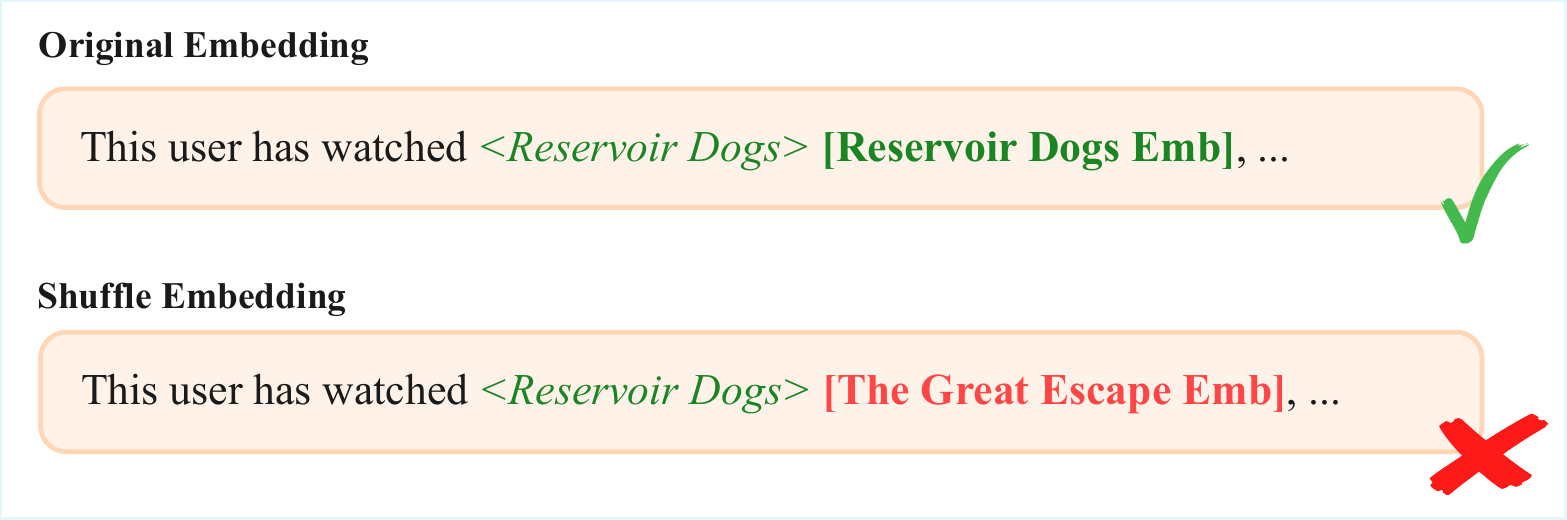}
    \end{minipage}
    \begin{minipage}[t]{1\columnwidth}
        \centering
        \small
        \setlength{\tabcolsep}{6pt}
        \begin{tabular*}{\linewidth}{@{\extracolsep{\fill}}lccc}
        \toprule
        \textbf{Model} & \textbf{HR@1 (Orig.)} & \textbf{HR@1 (Shuf.)} & $\boldsymbol{\Delta}$ \\
        \midrule
        Direct & 0.4531 & 0.4542 & $+0.0011$ \\
        LLaRA~\cite{liao2024llara}  & 0.4743 & 0.4678 & $-0.0065$ \\
        \bottomrule
        \end{tabular*}
    \end{minipage}

    \caption{\textbf{Embedding-shuffle diagnostic.}
    Shuffling collaborative embeddings while keeping titles fixed leaves MovieLens HR@1 (Hit Ratio at 1) nearly unchanged (\(\Delta=\) Shuf.\(-\)Orig. \(\approx 0\)),
    revealing weak reliance on collaborative embeddings.}
    \Description{Illustration of an embedding-shuffle diagnostic together with MovieLens HR@1 results. The item title remains unchanged while its collaborative embedding is replaced by another item's embedding. Direct changes from 0.4531 to 0.4542 HR@1 and LLaRA changes from 0.4743 to 0.4678, showing that shuffling collaborative embeddings has little effect on recommendation performance.}
    \label{fig:modality-bias}
\end{figure}

Despite their promising performance, these methods largely assume that once collaborative embeddings are mapped into the LLM embedding space, the model can effectively utilize the collaborative signals encoded in them. However, this assumption is far from guaranteed. A central challenge is the \textit{modality bias} between the two input modalities in LLM-based recommendation: textual item information and projected collaborative embeddings. Specifically, LLMs may over-rely on the textual signals while failing to effectively utilize the collaborative signals encoded in projected embeddings. This bias arises from a fundamental mismatch in representational structure. Item titles follow linguistic patterns that are well aligned with LLM pretraining, whereas user behavior sequences encompass diverse and often cross-domain transitions that are not naturally aligned with the LLM representation space. A single item may be followed by a wide range of subsequent items depending on context, intent, or temporal preferences. Consequently, directly projecting collaborative representations into the LLM embedding space does not guarantee that their underlying structure will be faithfully preserved or effectively utilized.

We verify this phenomenon quantitatively through an embedding-shuffle diagnostic. Using the MovieLens dataset~\cite{harper2015movielens}, we find that shuffling collaborative embeddings while preserving titles has a negligible impact on Hit Ratio at 1 (HR@1) for baselines that primarily rely on textual signals (Fig.~\ref{fig:modality-bias}). This suggests that projected collaborative embeddings may act more like generic soft prompts than true collaborative signals \citep{liu-etal-2022-p}, reflecting a stronger reliance on text than on collaborative information. This behavior mirrors findings in vision-language models where models remain confident even when non-textual inputs are corrupted \citep{lee2025vlindbench, zheng2025mllms}.

We attribute this bias to \textit{projection-induced distortion of collaborative structure}. Without explicit constraints, the projection module may fail to preserve the relational geometry originally captured by the sequential recommender, such as item-item similarities and sequence-aware relationships. To mitigate this issue, we propose a \textit{structure-preserving projection approach} that explicitly preserves collaborative structure in the projected LLM embedding space. Specifically, we introduce two objectives: (1) \textit{Cosine Similarity Preservation}, which aligns relational geometry between the original and projected spaces, and (2) \textit{Contrastive Preservation}, which preserves sequence-aware discrimination between positive and negative items.

Our contributions are threefold: (1) We provide quantitative evidence from the embedding shuffle tests that LLM-based recommenders may overlook the collaborative information encoded in projected embeddings. (2) We introduce a structure-preserving projection approach, enforced by two novel learning objectives designed to preserve item-item relational geometry and sequence-aware discrimination. (3) Through experiments on LastFM and MovieLens, we demonstrate consistent improvements in HR@1, and provide diagnostic analyses showing that our approach more effectively leverages collaborative signals.

\section{Methodology}
In this section, we introduce a structure-preserving projection approach that integrates collaborative representations into a pretrained LLM while explicitly preserving the relational structure of the original embedding space. The approach consists of: (1) a pretrained sequential recommender that provides collaborative embeddings, (2) an MLP projector that maps these embeddings into the LLM embedding space, and (3) two structure-preserving objectives that constrain the projection to maintain relational geometry.

\subsection{Model Overview}
Following recent LLM-based recommendation methods such as LLaRA and A-LLMRec, we formulate next-item prediction as a next-token generation task.
Given a user interaction sequence
$\mathcal{I} = [i_1, i_2, \dots, i_t]$, the goal is to predict the next item $i_{t+1}$.

Each item $i \in \mathcal{I}$ has a collaborative embedding $\mathbf{e}_i \in \mathbb{R}^d$ from a pretrained sequential recommender and a title with token embeddings $\mathbf{E}_{t_i} $. The collaborative item embeddings are projected by an MLP into the LLM embedding space $\mathbf{z}_i = \mathrm{MLP}(\mathbf{e}_i)$.

For each interaction, we insert both the title embedding and projected collaborative embedding into the LLM prompt:
\begin{equation}
    \mathbf{X} = \{\mathbf{E}_{\text{instruction}},\ \mathbf{E}_{t_{i_1}},\ \mathbf{z}_{i_1},\ \dots,\ \mathbf{E}_{t_{i_t}},\ \mathbf{z}_{i_t}\}.
\end{equation}

The LLM jointly leverages textual content and collaborative signals, trained with next-token prediction to minimize the negative log-likelihood of the ground-truth title tokens \(Y = \left[w_1, \ldots, w_m\right]\):
\begin{equation}
    \mathcal{L}_{\text{LM}} = - \sum_{j=1}^{m} \log P_\theta \left( w_j \mid w_{<j}, X \right).
\end{equation}

\subsection{Structure-Preserving Objectives}
To ensure that collaborative information is preserved in the projected space, we introduce two structure-preserving objectives:

\paragraph{(1) Cosine Similarity Preservation}
The first objective preserves the relational geometry of the original embedding space. Given item embeddings $\mathbf{e}_i$, $\mathbf{e}_j$ and their projected counterparts $\mathbf{z}_i=\mathrm{MLP}(\mathbf{e}_i)$, $\mathbf{z}_j=\mathrm{MLP}(\mathbf{e}_j)$, we define
$S_{ij}^{\text{orig}} = \cos(\mathbf{e}_i, \mathbf{e}_j)$ and $S_{ij}^{\text{proj}} = \cos(\mathbf{z}_i, \mathbf{z}_j)$. The loss is then computed as the mean squared difference between the two similarity matrices:
\begin{equation}
\mathcal{L}_{\text{sim}} = \frac{1}{|\mathcal{B}|^2} \sum_{i,j \in \mathcal{B}} \left(S_{ij}^{\text{proj}} - S_{ij}^{\text{orig}}\right)^2.
\end{equation}
where \(\mathcal{B}\) is the batch of item embeddings. 
This loss encourages the projector to preserve pairwise relational geometry by making item-item similarities in the projected space consistent with those in the original collaborative space.

\paragraph{(2) Contrastive Preservation}
The second objective ensures that the projection preserves the sequence-aware discrimination learned by the sequential recommender. Sequential models encode a preference ordering by producing higher similarity between a sequence representation and its true next item than with randomly sampled negative items. To preserve this behavior after projection, we use a contrastive objective that pulls the projected sequence embedding to align more closely with the projected positive item embedding, while pushing it away from projected negative item embeddings.

Given a sequence representation $\mathbf{h}_s \in \mathbb{R}^d$, the true next item embedding $\mathbf{e}_p$ and a randomly sampled negative embedding $\mathbf{e}_n$, $\mathbf{z}_s = \mathrm{MLP}(\mathbf{h}_s)$, $\mathbf{z}_p = \mathrm{MLP}(\mathbf{e}_p)$, and $\mathbf{z}_n = \mathrm{MLP}(\mathbf{e}_n)$. We optimize the objective using a binary cross-entropy loss:
\begin{equation}
\mathcal{L}_{\text{contrast}} = -\log \sigma\left( \langle \mathbf{z}_s, \mathbf{z}_p \rangle \right) - \log \big( 1 - \sigma( \langle \mathbf{z}_s, \mathbf{z}_n \rangle ) \big).
\end{equation}
The contrastive objective encourages projected representations to preserve the preference ordering learned by the sequential encoder. Specifically, it encourages projected sequence representations to have higher similarity with positive items and lower similarity with randomly sampled negative items.

\subsection{Optimization and Training}

The total training objective combines the next-token prediction loss with one of the proposed structure-preserving objectives:
\begin{itemize}
    \item \textbf{Cosine-LM}: 
    $\mathcal{L} = \mathcal{L}_{\text{LM}} + \alpha \mathcal{L}_{\text{sim}}$.

    \item \textbf{Contrastive-LM}: 
    $\mathcal{L} = \mathcal{L}_{\text{LM}} + \alpha \mathcal{L}_{\text{contrast}}$.
\end{itemize}
where $\alpha$ controls the strength of the structure-preserving objective.

Training proceeds in two stages. First, we pretrain the MLP projector using only the corresponding structure-preserving loss while keeping the sequential recommender and LLM fixed. This provides the projector with a structure-aware initialization before integrating the projected embeddings into the LLM. We then jointly optimize the projector and the trainable LLM parameters using the corresponding objective above, while keeping the sequential recommender frozen. The structure-preserving loss continues to constrain the projected representations during fine-tuning, preventing the collaborative structure from being excessively distorted by the language-modeling objective.

\section{Experiments}

We conduct experiments on two public datasets to evaluate the proposed approach and provide additional analyses to better understand its behavior and effectiveness.

\subsection{Experimental Setup}

\noindentparagraph{Datasets.}
We adopt the preprocessed LastFM~\cite{cantador2011hetrec} and MovieLens-100K~\cite{harper2015movielens} datasets released by LLaRA~\cite{liao2024llara}. Both datasets have also been widely used in recent LLM-based sequential recommendation studies, including iLoRA~\cite{kong2024ilora}, AFL~\cite{cai2025afl}, and S-DPO~\cite{chen2024softmaxdpo}. User interactions are chronologically ordered and split into training, validation, and test sets with an 8:1:1 ratio to prevent data leakage. Dataset statistics are summarized in Table~\ref{tab:dataset_statistics}.

\noindentparagraph{Baselines.}
We compare three categories of methods. (1) \emph{Traditional sequential models}: \textbf{GRU4Rec} and \textbf{SASRec}, which model user interaction sequences directly and also serve as collaborative embedding backbones. (2) \emph{Text-based LLM recommendation}: \textbf{LLaMA2-7B}~\cite{touvron2023llama} in a zero-shot setting and \textbf{TALLRec}~\cite{bao2023tallrec}, which instruction-tunes LLaMA2-7B for next-item prediction. (3) \emph{LLM recommendation with projected collaborative embeddings}: \textbf{Direct}, which jointly trains the projector and LLM using next-token prediction; \textbf{LLaRA}~\cite{liao2024llara}, which augments Direct with curriculum prompt tuning; and our \textbf{Cosine-LM} and \textbf{Contrastive-LM}, which additionally impose structure-preserving objectives on the collaborative representations.

\noindentparagraph{Implementation Details.}
We use the pretrained SASRec model released by LLaRA. GRU4Rec and all LLM-based methods follow the training setup of LLaRA. We fine-tune LLaMA2-7B with LoRA using cosine learning-rate scheduling and a batch size of 128. For the proposed structure-preserving losses, the coefficient $\alpha$ is selected on the validation set through a linear search with a step size of 0.1. The selected values are $\alpha=0.4$ for Cosine-LM and $\alpha=0.8$ for Contrastive-LM.

\begin{table}[t]
\centering
\small
\setlength{\tabcolsep}{8pt}
\begin{tabular}{lrrr}
\toprule
\textbf{Dataset} & \textbf{\# Users} & \textbf{\# Items} & \textbf{\# Interactions} \\
\midrule
LastFM & 1,220 & 4,606 & 73,510 \\
MovieLens-100K & 943 & 1,682 & 100,000 \\
\bottomrule
\end{tabular}
\caption{Statistics of the datasets.}
\label{tab:dataset_statistics}
\end{table}

\subsection{Evaluation Protocol}

\noindentparagraph{Candidate Construction.}
Following the evaluation setting of LLaRA, each test instance is associated with a candidate set of 20 items, consisting of the ground-truth next item and 19 sampled negative items. The same candidate set is used when evaluating ID-based and LLM-based methods, ensuring that all methods are compared under the same candidate space.

\noindentparagraph{Evaluation Procedure.}
For each test instance, all methods are evaluated on the same 20-item candidate set. GRU4Rec and SASRec first produce scores over the full item vocabulary, after which ranking is restricted to the candidate set. LLM-based methods instead generate an item title, which is normalized and matched against the candidate titles. A generation is considered valid only when it uniquely matches one candidate; unmatched or ambiguous generations are treated as invalid. A valid prediction is correct if the matched candidate corresponds to the ground-truth item.

\noindentparagraph{Metrics.}
Following LLaRA and S-DPO~\cite{chen2024softmaxdpo}, we report Hit Ratio at 1 (HR@1) and Valid Ratio (VR). For LLM-based methods, HR@1 is computed over valid generations as
$\mathrm{HR@1}=N_{\mathrm{correct}}/N_{\mathrm{valid}}$,
where $N_{\mathrm{correct}}$ is the number of valid predictions matching the ground-truth item and $N_{\mathrm{valid}}$ is the number of generations that uniquely match one candidate item. VR is defined as
$\mathrm{VR}=N_{\mathrm{valid}}/N_{\mathrm{test}}$,
where $N_{\mathrm{test}}$ denotes the total number of test instances. Invalid generations therefore reduce VR but are excluded from the denominator of HR@1. For ID-based methods, whose outputs always correspond to valid candidate items, VR is 1 and HR@1 is the fraction of test instances for which the ground-truth item is ranked first among the 20 candidates.

\begin{table}[t]
    \centering
    \small
    \setlength{\tabcolsep}{6pt}
    \begin{tabular}{lcc|cc}
        \toprule
        \textbf{Method} & \multicolumn{2}{c|}{\textbf{LastFM}$^*$} & \multicolumn{2}{c}{\textbf{MovieLens}$^\dagger$} \\
        \textbf{Metric} & HR@1 & VR & HR@1 & VR \\
        \midrule
        \multicolumn{5}{l}{\textit{Baseline}} \\
        \midrule
        GRU4Rec \citep{hidasi2015session} & 0.2616 & 1.0000 & 0.3750 & 1.0000 \\
        SASRec \citep{kang2018self} & 0.2233 & 1.0000 & 0.3444 & 1.0000 \\
        LLaMA2-7B \citep{touvron2023llama} & 0.0246 & 0.3443 & 0.0421 & 0.4421 \\
        TALLRec \citep{bao2023tallrec} & 0.4180 & 0.9836 & 0.3895 & 0.9263 \\
        \midrule
        \multicolumn{5}{l}{\textit{GRU4Rec Embedding}} \\
        \midrule
        Direct & 0.4872 & 0.9590 & 0.4043 & 0.9895 \\
        LLaRA \citep{liao2024llara} & 0.4344 & 0.9836 & 0.4421 & 0.9684 \\
        \textbf{Cosine-LM (ours)} & \underline{0.5133} & 0.9262 & \underline{0.4681} & 0.9895 \\
        \textbf{Contrastive-LM (ours)} & \textbf{0.5546} & 0.9754 & \textbf{0.4881} & 0.8842 \\
        \midrule
        \multicolumn{5}{l}{\textit{SASRec Embedding}} \\
        \midrule
        Direct & 0.4721 & 0.9967 & 0.4531 & 0.9789 \\
        LLaRA \citep{liao2024llara} & 0.4676 & 0.9820 & 0.4743 & 0.9811 \\
        \textbf{Cosine-LM (ours)} & \underline{0.5309} & 0.9885 & \underline{0.4874} & 0.9853 \\
        \textbf{Contrastive-LM (ours)} & \textbf{0.5511} & 0.9902 & \textbf{0.4993} & 0.9832 \\
        \bottomrule
    \end{tabular}
    \caption{Overall recommendation performance. Significance is tested on HR@1 via paired t-tests across five seeds against the strongest baseline per block: $^{*}$ denotes $p<0.05$ on LastFM, and $^{\dagger}$ denotes marginal significance at $p<0.1$ on MovieLens.}
    \label{tab:main_results}
\end{table}

\subsection{Overall Recommendation Performance}

As shown in Table~\ref{tab:main_results}, traditional sequential models (GRU4Rec, SASRec) markedly outperform zero-shot LLaMA2-7B, reflecting the limitations of text-only LLMs in leveraging collaborative signals. Through instruction tuning, TALLRec bridges this gap and surpasses both methods, confirming that targeted alignment can unlock the recommendation capabilities of LLMs.

Projected collaborative embeddings consistently improve performance over text-only baselines, indicating that collaborative signals provide complementary information beyond textual prompts. Our Cosine-LM and Contrastive-LM outperform Direct and LLaRA, with Contrastive-LM achieving the highest HR@1 on both datasets, indicating that contrastive alignment provides greater gains than cosine preservation. We assess statistical significance using paired t-tests across five random seeds. Our improvements are statistically significant on LastFM ($p<0.05$). On MovieLens, due to the exploratory nature of the analysis and higher variance across runs, we report marginal significance at $p<0.1$ (denoted by $^{\dagger}$).

\subsection{Analysis of Modality Reliance}
\label{sec:shuffle_analysis}

To verify whether the LLM truly leverages the collaborative signals from the projected embeddings or merely relies on the textual information from item titles, we conduct a shuffle-based ablation study. We introduce two noise settings during inference:

\textbf{Text Shuffle:} For each item in the interaction history, we replace its title with the title of a randomly sampled item from the global item pool, while keeping its original collaborative embedding unchanged. This breaks the correspondence between the interaction history and textual signals while preserving the collaborative input. A performance drop therefore indicates reliance on textual information.

\textbf{Embedding Shuffle:} For each item in the interaction history, we replace its collaborative embedding with that of a randomly sampled item from the global item pool, while keeping its original title unchanged. This corrupts the collaborative signals while preserving the textual input. A performance drop therefore indicates reliance on collaborative information.

\begin{figure}[t]
    \centering
    \begin{subfigure}{0.49\linewidth}
        \centering
        \includegraphics[width=\linewidth]{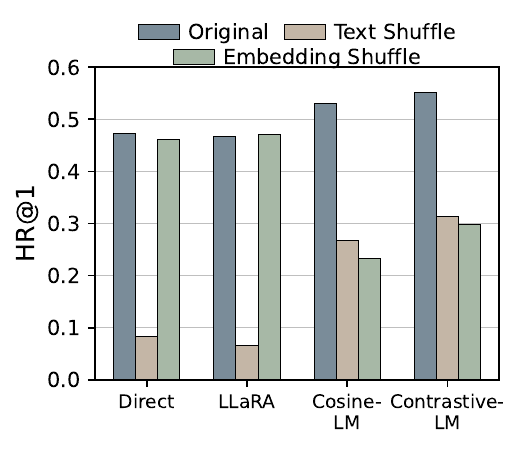}
        \caption{LastFM}
        \label{fig:shuffle_lastfm}
    \end{subfigure}
    \hfill
    \begin{subfigure}{0.49\linewidth}
        \centering
        \includegraphics[width=\linewidth]{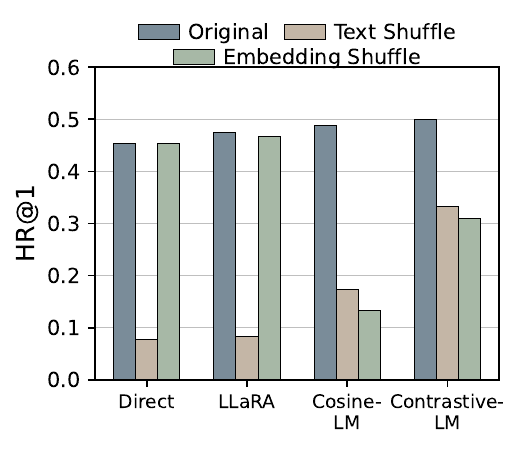}
        \caption{MovieLens}
        \label{fig:shuffle_movielens}
    \end{subfigure}
    \caption{Shuffle-based diagnostic of modality bias: baselines mainly rely on titles, whereas our methods actively leverage collaborative embedding structure.}
    \Description{Bar charts comparing HR@1 under the original, text-shuffled, and embedding-shuffled settings on LastFM and MovieLens for Direct, LLaRA, Cosine-LM, and Contrastive-LM. Text shuffling causes large performance drops for all methods. Embedding shuffling has little effect on Direct and LLaRA, but causes substantial drops for Cosine-LM and Contrastive-LM, indicating that the proposed methods rely more strongly on collaborative embeddings.}
    \label{fig:shuffle_results}
\end{figure}

Figure~\ref{fig:shuffle_results} reports HR@1 under the original setting and the two shuffle settings on LastFM and MovieLens, revealing a clear contrast between baseline and our structure-preserving approaches.

\emph{Over-reliance on Text in Baselines:}
For Direct and LLaRA, shuffling the text causes a catastrophic performance drop. By contrast, shuffling the embeddings results in negligible impact. This suggests that these baselines primarily function as text-driven recommenders: the projected embeddings are not effectively utilized as collaborative signals and are instead treated as generic soft prompts.

\emph{Effective Alignment in Proposed Methods:}
In contrast, our Cosine-LM and Contrastive-LM show substantial performance degradation when embeddings are shuffled. This sensitivity confirms that our structure-preserving objectives successfully encourage the LLM to utilize the geometric information encoded in collaborative embeddings. By enforcing structural alignment, our methods leverage both textual and collaborative signals and mitigate modality bias.

\begin{table}[t]
\centering
\resizebox{\columnwidth}{!}{
    \begin{tabular}{lccc|ccc}
    \toprule
    \textbf{Method} & \multicolumn{3}{c|}{\textbf{LastFM}} & \multicolumn{3}{c}{\textbf{MovieLens}} \\
    \textbf{Metric}       & Tau & Rho & SIM@10 & Tau & Rho & SIM@10 \\
    \midrule
    Direct & 0.2864 & 0.4151 & 0.2150 & 0.2334 & 0.3447 & 0.1237 \\
    LLaRA~\cite{liao2024llara}   & 0.3723 & 0.5290 & 0.2618 & 0.2064 & 0.3015 & 0.0879 \\
    \textbf{Cosine-LM} & \textbf{0.7801} & \textbf{0.9324} & \textbf{0.5327} & \textbf{0.7501} & \textbf{0.8573} & \textbf{0.5940} \\
    \textbf{Contrastive-LM} & \underline{0.6147} & \underline{0.8047} & \underline{0.4383} & \underline{0.3682} & \underline{0.5188} & \underline{0.2104} \\
    \bottomrule
    \end{tabular}
}
\caption{Structure preservation between original and projected item embeddings. Higher is better.}
\label{tab:structure_results}
\end{table}

\subsection{Embedding Structure Preservation}
\label{sec:structure-preservation}

We evaluate structure preservation by comparing the relational structure of the original collaborative space and the projected space using Kendall’s Tau, Spearman’s Rho, and SIM@10. For each item, all other items are ranked by cosine similarity in both spaces. We then compute Kendall’s Tau and Spearman’s Rho between the two rankings for each item and report the average across all items. SIM@10 is defined as the average overlap between the top-10 nearest neighbors in the original and projected spaces. While Tau and Rho measure global ranking consistency across the entire item space, SIM@10 focuses on local neighborhood preservation by examining whether the nearest neighbors of each item remain stable after projection.

Table~\ref{tab:structure_results} shows that Cosine-LM achieves substantially stronger geometric preservation, consistent with its explicit objective of matching pairwise similarities in the original collaborative space. Interestingly, however, this stronger geometric fidelity does not translate directly into the best recommendation performance: Contrastive-LM achieves higher HR@1 despite lower Tau, Rho, and SIM@10. This difference highlights an important trade-off between \emph{global geometric preservation} and \emph{task-specific discriminative preservation}. Cosine-LM attempts to retain the overall item–item neighborhood structure, including relationships that may not be directly relevant to next-item prediction. In contrast, Contrastive-LM is free to reshape parts of this geometry as long as sequence representations remain better aligned with positive next items than with negative items. Thus, preserving the original embedding space more faithfully is not necessarily equivalent to preserving the information most useful for recommendation. These results suggest that moderate structural distortion can be beneficial when it improves task-relevant discrimination.

\subsection{Synergy of Combined Loss}
\label{sec:synergy}

While Cosine-LM excels at preserving geometric structure and Contrastive-LM provides superior discriminative power, we hypothesize that these objectives are complementary. Initiating training with geometric constraints can prevent early structural distortion, establishing a stable topology that facilitates the subsequent learning of sequence-aware discrimination.

\begin{figure}[t]
    \centering
    \begin{subfigure}[b]{0.49\columnwidth}
        \centering
        \includegraphics[width=\linewidth]{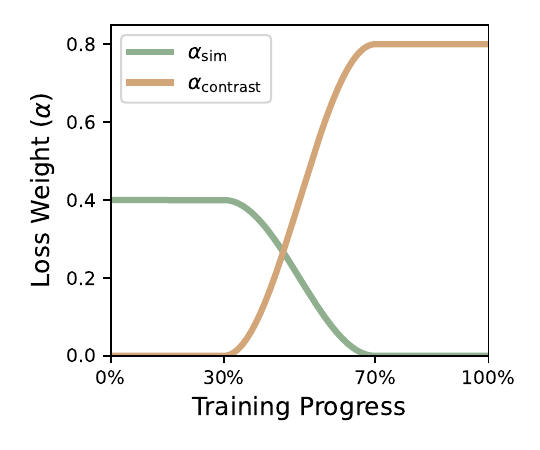}
        \caption{Dynamic Weight Schedule}
        \label{fig:loss_weights}
    \end{subfigure}
    \hfill
    \begin{subfigure}[b]{0.50\columnwidth}
        \centering
        \includegraphics[width=\linewidth]{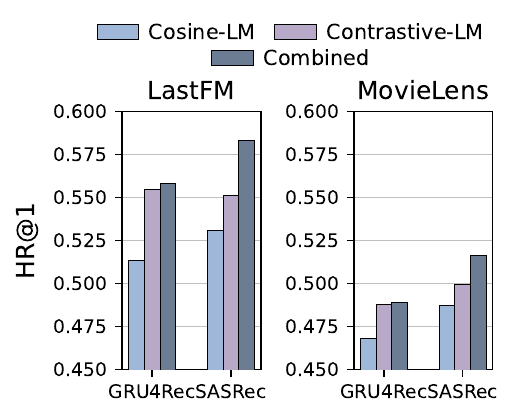}
        \caption{Performance Comparison}
        \label{fig:hr1_both}
    \end{subfigure}
    \caption{\textbf{Synergy Analysis.} (a) The dynamic schedule shifts from geometric preservation to contrastive alignment using cosine interpolation. (b) Combined yields the best HR@1 on both datasets, confirming the effectiveness of the curriculum.}
    \Description{Two-panel synergy analysis. The left panel shows the training schedule: the cosine-similarity loss weight starts at 0.4, decreases between 30 and 70 percent of training, while the contrastive-loss weight increases to 0.8 and remains there. The right panel compares HR@1 for Cosine-LM, Contrastive-LM, and Combined using GRU4Rec and SASRec embeddings on LastFM and MovieLens. Combined achieves the highest HR@1 in all shown settings.}
    \label{fig:synergy_analysis}
\end{figure}

To verify this hypothesis, we adopt a three-phase training schedule that gradually shifts the optimization focus from geometric preservation to task-specific discrimination. During the \textbf{Warm-up} phase (0\%–30\%), we set $\alpha_{\text{sim}}=0.4$ and optimize primarily for cosine similarity preservation, encouraging the projector to establish a stable mapping that retains the relational structure of the original collaborative space. During the \textbf{Transition} phase (30\%–70\%), $\alpha_{\text{sim}}$ is gradually decayed to 0 while $\alpha_{\text{contrast}}$ is increased to 0.8 using cosine interpolation, allowing the model to progressively trade global geometric fidelity for sequence-aware discrimination without an abrupt change in the optimization objective. Finally, in the \textbf{Refinement} phase (70\%–100\%), we keep $\alpha_{\text{contrast}}=0.8$, enabling the model to focus on distinguishing positive next items from negatives after a structurally meaningful projection has already been established. Figure~\ref{fig:loss_weights} illustrates this schedule.

As shown in Figure~\ref{fig:hr1_both}, \textbf{Combined} consistently outperforms both individual objectives on LastFM and MovieLens. This result supports the complementary roles of the two losses: cosine preservation provides a well-structured initialization that limits early distortion of collaborative relationships, while contrastive learning subsequently reshapes this structure toward the recommendation objective. The resulting curriculum therefore balances geometric fidelity and task-specific discrimination, leading to better recommendation performance than optimizing either objective alone.

\begin{table}[t]
\centering
\renewcommand{\arraystretch}{0.9}
\resizebox{\columnwidth}{!}{
    \begin{tabular}{lcc|cc}
        \toprule
         & \multicolumn{2}{c|}{\textbf{LastFM (HR@1)}} & \multicolumn{2}{c}{\textbf{MovieLens (HR@1)}} \\
         {\textbf{Method}} & Original & Fine-tuned & Original & Fine-tuned \\
        \midrule
        Direct & 0.4721 & 0.4852 & 0.4531 & 0.4574 \\
        LLaRA~\cite{liao2024llara}  & 0.4676 & 0.4754 & 0.4743 & 0.4787 \\
        \textbf{Cosine-LM} & \underline{0.5309} & \underline{0.5492} & \underline{0.4874} & \underline{0.5026} \\
        \textbf{Contrastive-LM} & \textbf{0.5511} & \textbf{0.5702} & \textbf{0.4993} & \textbf{0.5106} \\
        \bottomrule
    \end{tabular}
}
    \caption{Ablation of joint fine-tuning. "Original" keeps SASRec frozen, while "Fine-tuned" updates SASRec jointly with LLM. Our methods remain consistently better than baselines.}
    \label{tab:fine-tuned_results}
\end{table}

\subsection{Ablation of Joint Fine-tuning}
\label{sec:joint_fine-tune}

To investigate the potential for further performance improvements, we conduct an ablation study on jointly fine-tuning the sequential backbone (SASRec) with the LLM and projector. In our main experiments, SASRec was kept frozen to isolate the projector’s alignment capability. Here, we unfreeze it to allow collaborative embeddings to adapt towards the LLM representation space.

Table~\ref{tab:fine-tuned_results} compares the frozen-backbone setting (Original) and the joint fine-tuning setting (Fine-tuned). While enabling backbone updates improves all methods, our structure-preserving approaches maintain a clear advantage over strong baselines under both settings. Notably, the performance gains from joint fine-tuning are larger for Cosine-LM and Contrastive-LM than for Direct and LLaRA, suggesting a synergistic effect between backbone adaptation and our geometry-aware objectives. When SASRec becomes trainable, it can actively reshape the collaborative embedding topology to better satisfy the geometric and contrastive constraints, thereby strengthening the manifold alignment with the LLM.

\section{Conclusion}

In this paper, we identify and address a previously overlooked phenomenon of modality bias in LLM-based recommenders that integrate textual and collaborative signals. We propose a structure-preserving projection approach with two complementary objectives: cosine similarity preservation and sequence-aware contrastive alignment, enabling more faithful integration of collaborative signals into LLMs. Experiments across two datasets and multiple collaborative backbones confirm the prevalence of modality bias and show that our approach effectively preserves collaborative information while substantially improving recommendation accuracy. These findings establish structure-preserving projection as an effective solution for reliable LLM-based recommendation. Future work includes expanding the method to richer modalities, evaluating it against a broader range of recommendation models, and developing interpretable mechanisms that explicitly surface collaborative cues.

\begin{acks}
We thank the anonymous reviewers for their insightful and constructive feedback.
This work was supported by the National Science and Technology Council (NSTC), Taiwan, under Grant NSTC 115-2634-F-001-006.
\end{acks}

\bibliographystyle{ACM-Reference-Format}
\bibliography{reference}

% \appendix

\end{document}